\documentclass[letterpaper, 11pt, onecolumn]{IEEEtran}
\usepackage[margin=1.0in]{geometry}
\usepackage[pdftex]{graphicx}
\usepackage{xcolor}
\usepackage{multirow}
\usepackage{algorithm}
\usepackage{balance}
\usepackage{placeins}
\usepackage{array}
\usepackage{blindtext}
\usepackage{subfigure}
\usepackage{caption}
\usepackage{capt-of}
\usepackage{booktabs}
\usepackage{amsmath,amssymb,amsfonts}
\usepackage{algorithmic}
\usepackage{graphicx}
\usepackage{textcomp}
\usepackage{supertabular}
\usepackage{lineno,hyperref}

\begin{document}

%

\title{Programmable Transimpedance Amplifier with Integrated Bandgap Reference for Glucose Concentration Measurement}

%

	

\author{
\begin{tabular}{ccc}
				\\
 Riyaz Ahmad &	 Amit M. Joshi   &Dharmendar Boolchandani  \\
Dept. of ECE &  Dept. of ECE  &
Dept. of ECE \\
MNIT, Jaipur, India  &  MNIT, Jaipur, India  & MNIT, Jaipur, India\\
2018rec9120@mnit.ac.in &  amjoshi.ece@mnit.ac.in   & dboolchandani.ece@mnit.ac.in
			\end{tabular}
}

\maketitle

\begin{abstract}
For glucose electrochemical sensors, a comprehensive electronics interface is designed and constructed in 0.18 $\mu$m, CMOS process technology, and 1.5 V supply voltage. This interface includes a programmable readout amplifier and bandgap reference voltage potentiostat circuit. The programmable transimpedance amplifier (PTIA), the proposed readout circuit, provides a large dynamic range and low noise. Overall transimpedance increase for the PTIA is 17.3-50.5 k$\Omega$. For an input current range of 4.2-180 $\mu$A, the PTIA response has a linear output voltage range of 0.55-1.44 V. The output rms noise value is calculated to be 5.101 $\mu$V{$_{\scriptsize rms}$} and the overall power consumption of the design is 2.33 mW. In the current range mentioned above, the THD percentage spans from 7.6 to 10.2. All measurements from the bandgap reference voltage potentiostat are made using the reference potential of 0.6 V. The working electrode was a glassy carbon electrode (GCE) loaded with a CuO/Cu{$_{\scriptsize 0.76}$}CO{$_{\scriptsize 2.25}$}O{$_{\scriptsize 4}$} (copper cobaltite) coating. Electrochemical glucose sensing setup has been used to measure glucose concentrations between 1 and 10 mM, and an emulated circuit has been used to verify the viability of the proposed glucose sensing design. The suggested glucose sensor architecture has a total size of 0.0684 mm{\textsuperscript 2}.  
\end{abstract}




\section{Introduction}
\label{sec1}
Chronic diabetes is characterised by high blood glucose levels in the body \cite{joshi2020iglu2.0,yang2021robust,jain2019iglu}. Diabetes, which has tripled in prevalence over the past 20 years due to an unbalanced glycemic profile \cite{basu2020use,joshi2021everything}. Diabetes is one of the critical health issues with the quickest rate of growth \cite{li2019canet,9203858,9687837,agrawal2022machine}. The body needs glucose to carry out daily functions, however the normal range of glucose can be considered to be between 4.4 mM and 8.33 mM, or between 80 $\frac{mg}{dl}$ to 150 $\frac{mg}{dl}$ \cite{jain2021everythingNew}, therefore it needs to be routinely monitored \cite{ahmad2021,noaro2020machine,ahmad2022,zhang2020noninvasive,ahmad2023novel}. 
Hyperglycemia, also known as a high blood glucose level, causes blood vessels to harden, which can damage kidneys, cause blindness, and sometime even cause the failure of the organs \cite{joshi2020smart}. Diabetes is linked to cardiac disease, peripheral vascular disease, and amputation of limbs. On the other hand, Type 1 Diabetes (T1DM) patients may experience low blood sugar, or hypoglycemia, due to an excessive insulin dosage. The most frequent hypoglycemia symptoms are 50\% more prevalent in persons with diabetes than in those without \cite{zhu2022personalized,jain2020iglu, cappon2023individualized}. Diabetes also adds to the cost of care at the delivery and treatment point. The presence of diabetes patients may also reduce job productivity and increase the risk of incapacity \cite{jain2021review}. Diabetes can also lead to a number of health problems, including depression, digestive troubles, anxiety disorders, mental disorders, and altered eating patterns. Diabetes could be managed with a healthy diet, some exercise, insulin dosage, and medications \cite{yau2023reinforcement,noaro2023personalized}. With oral medications, diabetes can be controlled in its early stages. Controlling diabetes also lowers the risk of high blood pressure, cardiovascular disease, and amputation. Dizziness, perspiration, and exhaustion are the symptoms, and in the worst case, they might cause a coma and death. The common signs of diabetes include thirst, fatigue, changes in vision, persistent hunger, sudden weight loss, and the frequent passing of urine in brief bursts \cite{jain2019iglu}.
Diabetes can result in blindness, heart attacks, kidney illness, lower limb amputations, and blindness if it is left untreated for an extended length of time. 
Diabetes can also lead to a number of health problems, including depression, digestive troubles, anxiety disorders, mental disorders, and altered eating patterns \cite{ramesh2021remote}.
Diabetes can be controlled with the right diet, occasional activity, insulin dosage, and medications. With oral medications, diabetes can be controlled in its early stages. Controlling diabetes also lowers the risk of high blood pressure, cardiovascular disease, and amputation \cite{joshi2020iglu}.

In order to measure body glucose, it is always necessary to use an electronic interface that can handle a wide range of different glucose concentrations with improved sensitivity and linearity. For processing the signals from electrochemical glucose sensors, a readout circuit and reference potentiostat-based interface architecture is useful. Three electrodes of electrochemical cells, also known as amperometric sensors, employed for glucose sensing are the Reference Electrode (RE), Working Electrode (WE), and Counter Electrode (CE) \cite{bard2001fundamentals}. The electrochemical estimations were performed using a three-electrode design arrangement, with Pt foil (1 $\times$ 2 cm) serving as the CE and Ag/AgCl serving as the RE. The WE is the CuO/Cu{$_{\scriptsize 0.76}$}CO{$_{\scriptsize 2.25}$}O{$_{\scriptsize 4}$} coated GCE filled.

In order to handle sensor signals, current mode (CM) circuits provide a number of advantages over their voltage mode (VM) equivalents, including a lower power requirement \cite{ferri2003low,ahmad2022,ahmad2023novel}. The majority of the structures created for the processing of electrochemical sensor signals use op-amps-based VM circuitry, while very little study has been done using CM circuitry. The literature review \cite{ghosal2021glucam,ying2021,jain2020iglu1.1} examines interface concepts for electrochemical sensing and systems for monitoring glucose level. In the \cite{lu2021review}, the applications of electrochemical sensor signal processing and features of instrument development were discussed. The WE was not connected to true ground, rendering it sensitive to external disturbances, and the transimpedance amplifier (TIA) \cite{busoni2002comparison} was employed to handle low sensor current with high feedback resistors. As a result, the circuit uses a lot of space and produces a lot of noise. A current-conveyor (CC) based TIA with subpar sensitivity for glucose sensing was reported in \cite{esparza2014,li2016}. The CM and resistive TIA based readout circuits were reportedly run in comparison at higher power dissipation, according to \cite{wang2010}. The TIA in \cite{park2022}, is digitally adjustable with a complicated digital circuit and covers a wide range of sensor current. The low range of current within $\pm$2 $\mu$A is measured by the readout circuit in \cite{li2022in}.

Current was detected using three op-amp-based instrumentation amplifiers in \cite{kuo2021novel} and an op-amp-based current mirror readout design in \cite{ahmadi2008current}. Although \cite{martin2009fully} gives a better potential control, it has some of shortcomings of \cite{busoni2002comparison}. An enhanced sensing technique was utilised in \cite{aleeva2018amperometric,hanitra2019multichannel}, although at the expense of a complex structure and multiple passive components. Karandikar et al. \cite{karandikar2016low} proposed a compact design that uses a high load resistance to detect small sensor currents. In \cite{shenoy2014,ghore2017}, a small range of glucose concentration with a larger power dissipation was discussed. According to \cite{ahmadi2009wireless,al2016,puttan2020} and \cite{yang2009}, the small range of sensor current detection requires a high power supply.

The majority of the topologies covered in the aforementioned literatures are found to be VM, or op-amp based, necessitating different external circuitry for the creation of the reference voltage and the readout circuit to convert sensor current into the proper output voltage \cite{tripathi2021design}. A design with externally generated reference voltage is more complex, uses more power, and produces a lot of input referred noise. 

The following are some novel contributions made by this work:

\begin{itemize}
	\item  For glucose sensing, a novel block level architecture has been proposed. 
	\item A programmable readout circuit and a bandgap reference voltage potentiostat are merged into a single circuit.
	\item  In a three electrode electrochemical sensing setup for measuring glucose concentration, the proposed design has a higher current processing capability than the current measured by the cyclic voltammetry (CV) method.
	\item Also, the proposed glucose sensing architecture emulated circuit has been developed to show the working feasibility of. 
\end{itemize}

Section \ref{sec2} explains the specific glucose sensing interface architecture as well as the CMOS implementation, design, and analysis of the suggested architecture, which includes PTIA and a bandgap reference voltage potentiostat circuit for glucose sensing. Section \ref{sec3} discusses the sensor's CV and CA electrochemical measurement details, proposed design performance findings, the setup of a simulated circuit and measured results, and comparisons with other works. The task is finally finished at section \ref{sec4}.

\section{Proposed architecture for glucose sensing applications}
\label{sec2}
\begin{figure}[!t]
	\centerline{\includegraphics[scale=0.6]{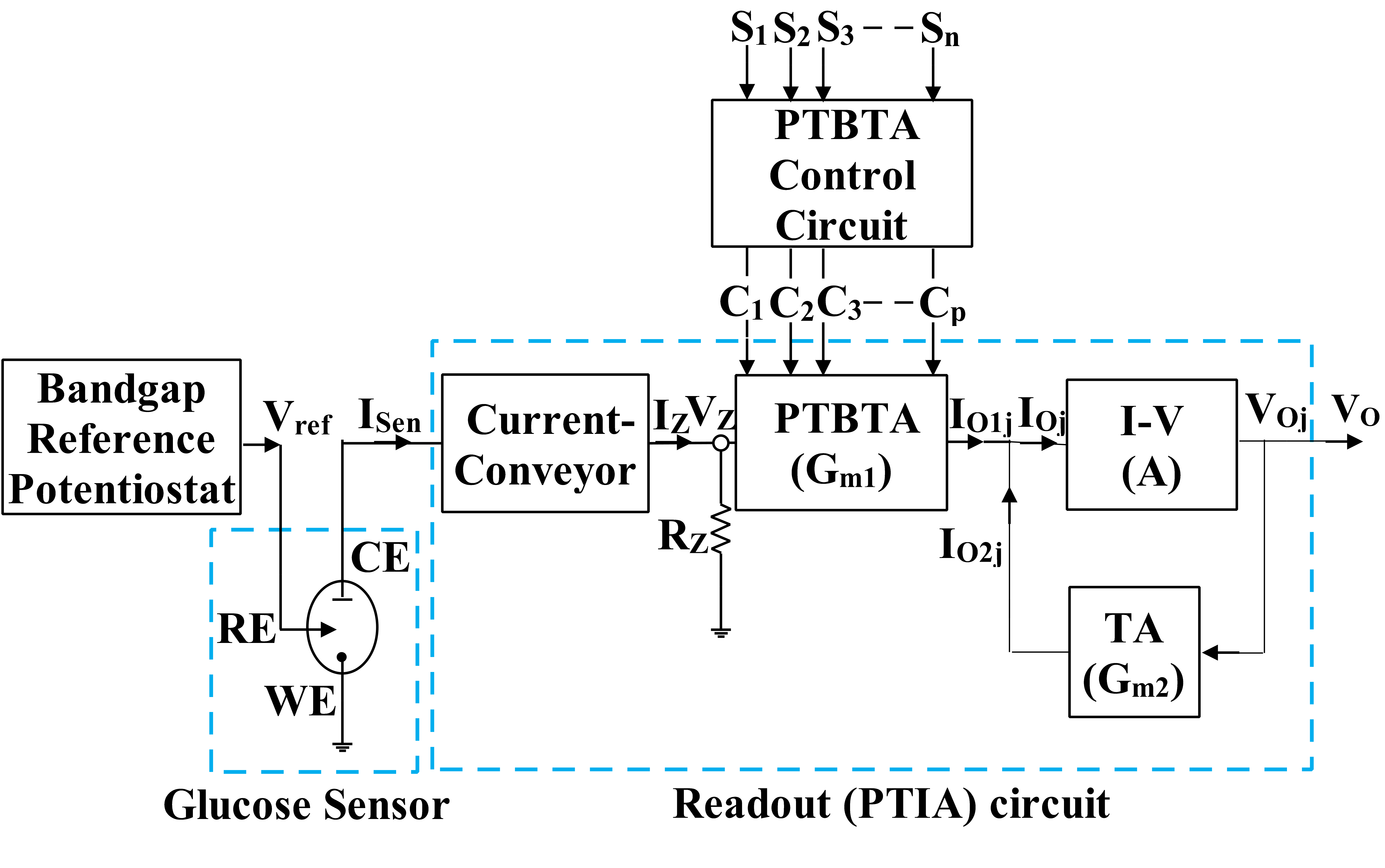}}
	\caption{Block architecture for glucose level monitoring.}
	\label{fig1}
\end{figure}

In the suggested electronics architectural block given in Fig. \ref{fig1} for glucose level measurement, a bandgap reference voltage potentiostat and programmable readout circuits are used to connect with electrochemical glucose sensors. An electrochemical sensor's WE and RE electrodes are coupled to a reference potentiostat. The CE terminal current, or sensor current, is supplied into a differential CC to efficiently buffer the sensor current due to CCs' several advantages over current mirror circuits. The PTBTA circuit amplifies the CC output current. The PTBTA control logic circuit is used to amplify the current at different levels, and the simple NMOS current to voltage (I-V) converter and TA circuit feedback arrangement convert the amplified current signal to voltage at the end.

Subsection \ref{sec2.1} has a detailed explanation of the suggested architecture, which consists of a bandgap reference voltage potentiostat, a PTIA circuit (readout) based on CC, programmable PTBTA, a current to voltage converter, and TA circuit for glucose sensing. Following that, a thermal noise analysis for PTIA has been covered in subsection \ref{sec2.2}.

\subsection{Bandgap reference voltage potentiostat and programmable readout circuit analysis in electronics architecture}
\label{sec2.1}

\begin{figure*}[!h]
	\centerline{\includegraphics[scale=0.65]{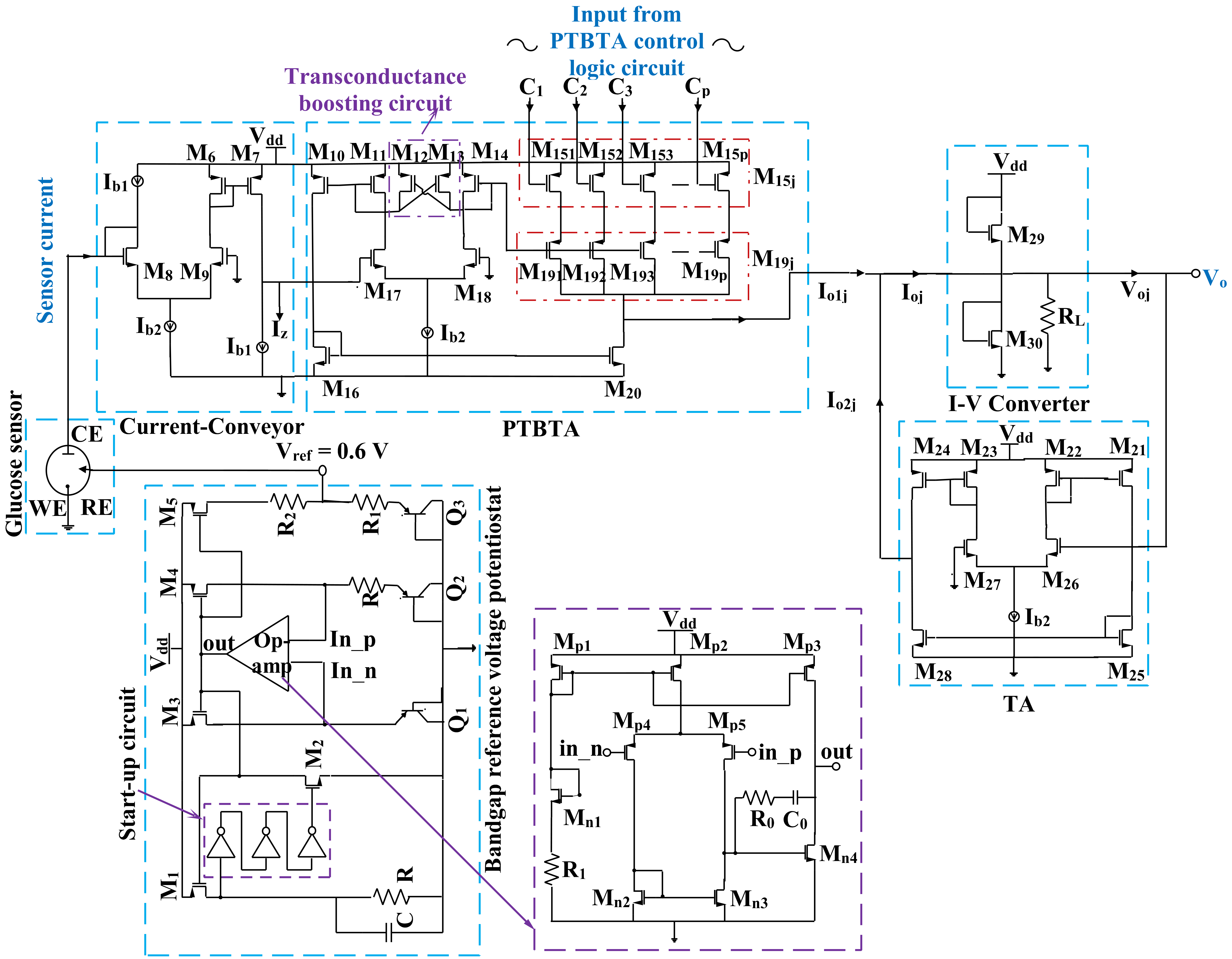}}
	\caption{Circuit diagram of electronics interface: Bandgap reference voltage potentiostat and programmable readout amplifier circuit.}
	\label{fig2}
\end{figure*}

CMOS based implementation has been considered for the implementation of the proposed design \cite{tripathi2018carbon}.
Fig. \ref{fig2} shows a fully functional CMOS implementation of the electronic interface for glucose sensing. In order to process the current flowing through the CE electrode, electrochemical sensors need a steady reference voltage to drive the electrode. This effort has produced a 0.6 V reference potential. The various components of a full architecture are also described and depicted in Fig \ref{fig2}.

The sensor current I{$_{\scriptsize sen}$} is made possible by the differential CC which produces current I{$_{\scriptsize z}$}
to have better processing and buffering. The relation between  I{$_{\scriptsize sen}$} and I{$_{\scriptsize z}$} is expressed as

\begin{equation}
I_{z}=\alpha_{c} I_{sen}
\label{eqn:1}
\end{equation}
Where, $\alpha${$_{\scriptsize c}$} is the current tracking error of differential CC. The ideal value of $\alpha${$_{\scriptsize c}$} is close to unity. So,

\begin{equation}
I_{z}\approx I_{sen}
\label{eqn:2}
\end{equation}
Cross-coupled MOS transistors M12 and M13 have been connected to increase the PTBTA's transconductance. By maintaining the transconductance ratio of transistors M12 and M11 close to unity, it is possible to increase the sensor current without increasing the bias current. The effective value of PTBTA transconductance G{$_{\scriptsize m1}$} can be calculated using the DC transfer expression \cite{ahmad2023}.

\begin{equation}
G_{m1}=\dfrac{g_{m17}}{1-\dfrac{g_{m12}}{g_{m11}}}
\label{eqn:3}
\end{equation}

The output signal in voltage form is obtained by applying the current signal from the PTBTA circuit to a feedback system made up of an I-V converter and TA, as shown in Fig. \ref{fig2}.

\begin{figure}[!h]
	\centerline{\includegraphics[scale=0.52]{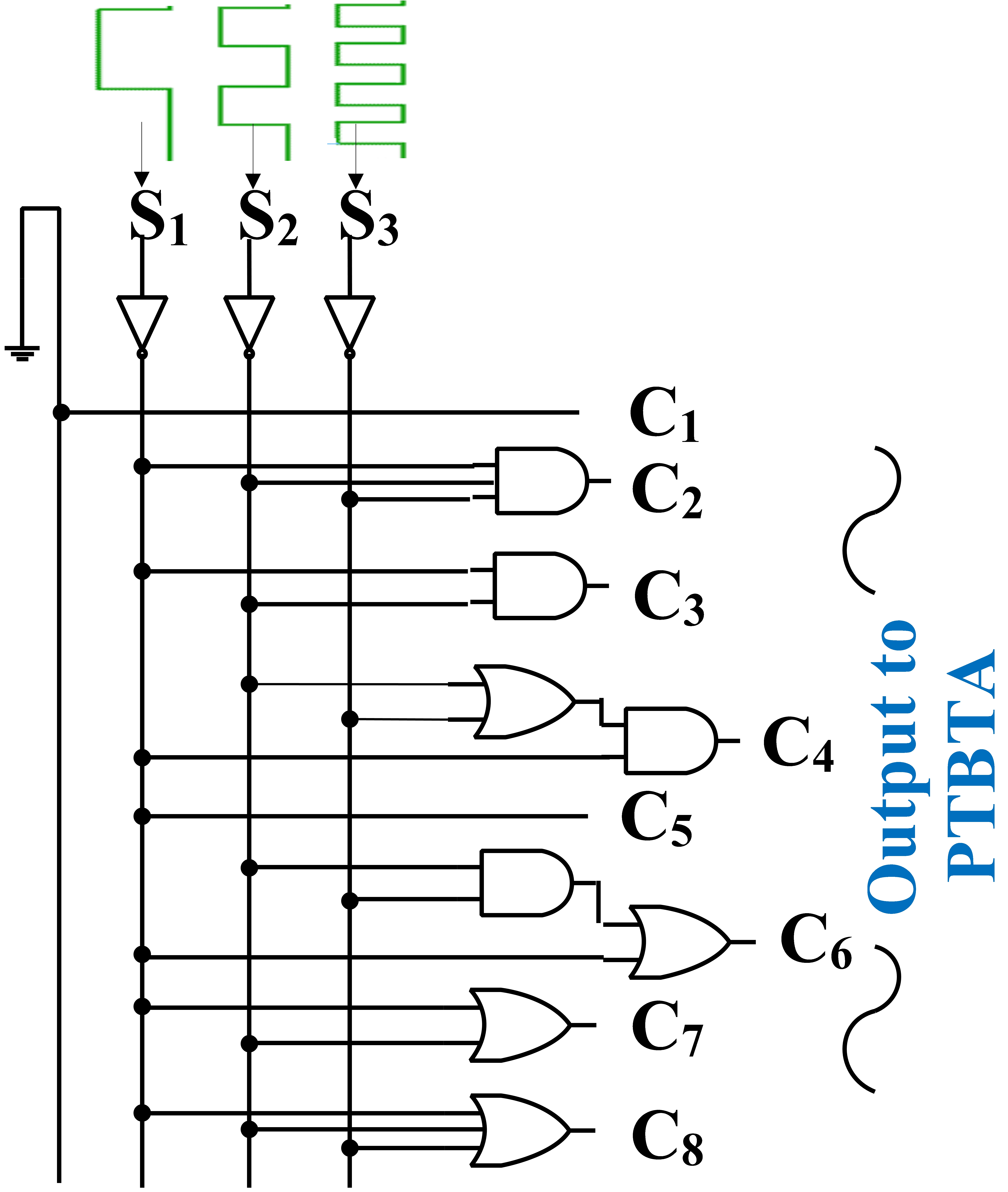}}
	\caption{Programmable transconductance boosted transconductance amplifier (PTBTA) control logic circuit.}
	\label{fig3}
\end{figure}

In Fig. \ref{fig3}, the PTBTA control logic circuit is depicted. Three select input lines (S{$_{\scriptsize 1}$}, S{$_{\scriptsize 2}$} and S{$_{\scriptsize 3}$}) are used to generate eight control signals (C{$_{\scriptsize 1}$}, C{$_{\scriptsize 2}$}.......C{$_{\scriptsize 8}$}). The definition of the truth table in Table \ref{tbl1} shows select lines and control signals.

\begin{table}[h!]
	\centering
	\renewcommand{\tabcolsep}{2mm}
	\caption{PTBTA control logic circuit's truth table}\label{tbl1}
	\begin{tabular}{ccc|cccccccc}
	\hline
		S{$_{\scriptsize 1}$} & S{$_{\scriptsize 2}$}& S{$_{\scriptsize 3}$} & C{$_{\scriptsize 1}$}&C{$_{\scriptsize 2}$}&C{$_{\scriptsize 3}$}&C{$_{\scriptsize 4}$}&C{$_{\scriptsize 5}$}&C{$_{\scriptsize 6}$}&C{$_{\scriptsize 7}$}&C{$_{\scriptsize 8}$} \\
		\hline
		0 & 0 & 0&0&1&1&1&1&1&1&1 \\
		0 & 0 & 1&0&0&1&1&1&1&1&1 \\
		0 & 1 & 0&0&0&0&1&1&1&1&1 \\
		0 & 1 & 1&0&0&0&0&1&1&1&1 \\
		1 & 0 & 0&0&0&0&0&0&1&1&1 \\
		1 & 0 & 1&0&0&0&0&0&0&1&1 \\
		1 & 1 & 0&0&0&0&0&0&0&0&1 \\
		1 & 1 & 1&0&0&0&0&0&0&0&0 \\
		\hline
	\end{tabular}
\end{table}

The following is the equation for the PTBTA's programmable output current:
\begin{equation}
I_{oj}=j I_{o}
\label{eqn:4}
\end{equation}
Where, j = 1, 2,....p, so the corresponding currents at output, I{$_{\scriptsize o1}$}, I{$_{\scriptsize o2}$}, I{$_{\scriptsize o3}$}.......I{$_{\scriptsize op}$} of PTBTA are equal to I{$_{\scriptsize o}$}, 2I{$_{\scriptsize o}$}, 3I{$_{\scriptsize o}$}, ..........,pI{$_{\scriptsize o}$}, respectively. Where, the output current I{$_{\scriptsize o}$} (i.e. for j = 1) of PTBTA is manifested as
\begin{equation}
I_{o}=G_{m1} V_{z}=G_{m1} R_{z}I_{z}
\label{eqn:5}
\end{equation}
Equation \ref{eqn:5} is changed to \ref{eqn:6} by setting the value of  I{$_{\scriptsize z}$} from Eq.\ref{eqn:1} into Eq.\ref{eqn:5}.

\begin{equation}
I_{o}=\alpha_{c} G_{m1} R_{z} I_{sen}
\label{eqn:6}
\end{equation}

The amount $\alpha_{c}$G{$_{\scriptsize m1}$}R{$_{\scriptsize z}$} reflects the PTBTA's current gain in \ref{eqn:6}. A straightforward I-V converter and TA circuit in feedback configuration are utilised to convert I{$_{\scriptsize o}$} to voltage, improving the stability and transimpedance gain of the overall programmable transimpedance amplifier (PTIA) circuit. The readout amplifier's transimpedance gain (PTIA) is finally stated as

\begin{equation}
\dfrac{V_{o}}{ I_{sen}}= \dfrac{\alpha_{c} G_{m1} R_{z} A}{(1-AG_{m2})}
\label{eqn:7}
\end{equation}
Where, A is gain of I-V converter and G{$_{\scriptsize m2}$} is transconductance of TA in the feedback system. By setting the value of AG{$_{\scriptsize m2}$} close to unity in Eq.\ref{eqn:7}, a large value of transimpedance gain can be achieved for the proposed PTIA circuit. The expression of A can be expressed as
\begin{equation}
A = \dfrac{1}{2k (V_{dd}-V_{th})} 
\label{eqn:8}
\end{equation}
Where, k and V{$_{\scriptsize th}$} are the process parameter and threshold voltage of NMOS transistor M29 and M30 of I-V converter. Additionally, the general formula for the PTIA output voltage is given as
\begin{equation}
V_{oj}= j V_{o}
\label{eqn:9}
\end{equation}

For j = 1, 2,...p, the corresponding output voltages V{$_{\scriptsize o1}$}, V{$_{\scriptsize o2}$}, V{$_{\scriptsize o3}$}......V{$_{\scriptsize op}$} of PTIA are equal to V{$_{\scriptsize o}$}, 2V{$_{\scriptsize o}$}, 3V{$_{\scriptsize o}$}, .......pV{$_{\scriptsize o}$}, respectively.

\subsection{Noise analysis}\label{sec2.2}
\begin{figure}[!h]
	\centerline{\includegraphics[scale=0.75]{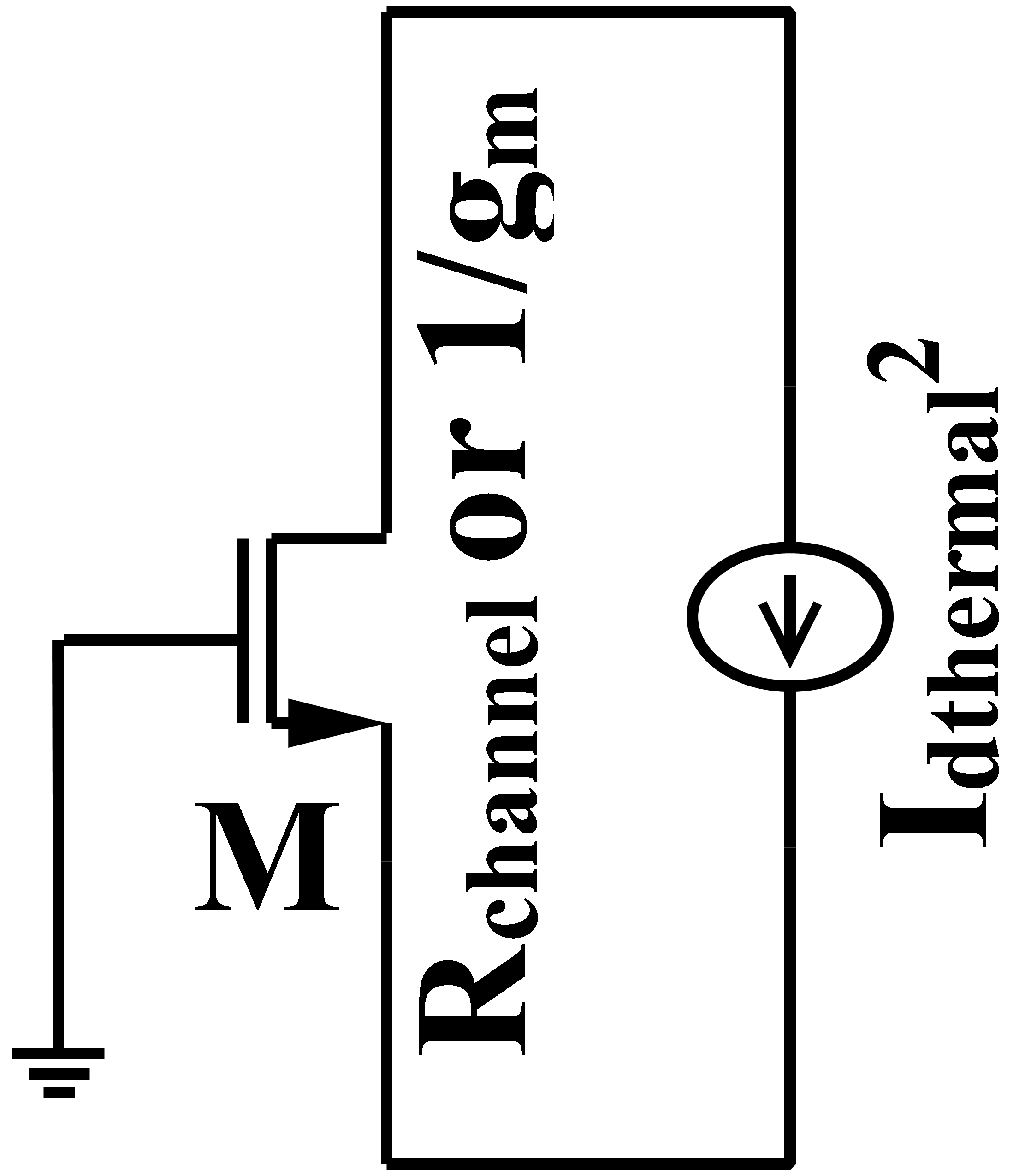}}
	\caption{MOS transistor thermal noise model.}
	\label{fig4}
\end{figure}

The thermal noise analysis is provided for the readout or PTIA circuit. Given in Eq.\ref{eqn:10} is the thermal noise current generated by the MOS transistor depicted in Fig. \ref{fig4}.

\begin{figure}[!htb]
	\centerline{\includegraphics[scale=0.62]{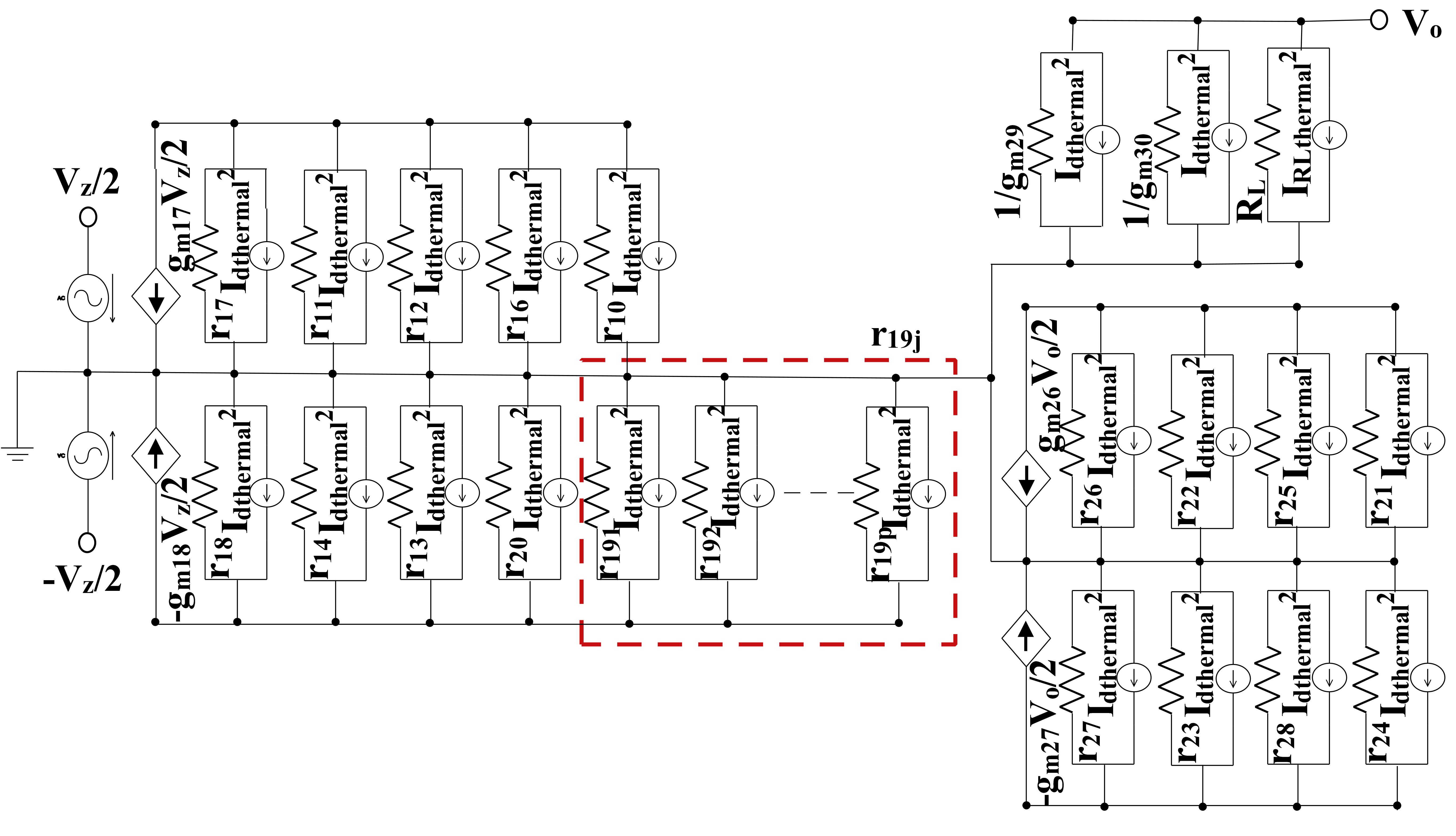}}
	\caption{Equivalent small signal noise model of readout or PTIA circuit.}
	\label{fig5}
\end{figure} 

\begin{equation}
\overline{I_{dthermal}{\textsuperscript 2}}= \dfrac{4 kT\Gamma} {R_{channel}} = 4 kT\Gamma g_{m}
\label{eqn:10}
\end{equation}
Where (R{$_{\scriptsize channel}$} = $\frac{1}{g_{\scriptsize m}}$). Fig. \ref{fig5} shows the corresponding noise model circuit for PTIA. The input devices g{$_{\scriptsize m}$}/I{$_{\scriptsize D}$} or g{$_{\scriptsize m}$} kept high \cite{binkley2007tradeoffs} to dominate the overall readout noise in general. Thermal noise's contribution from the PTIA has been studied. The proposed equation for the power spectral density (PSD) of the PTIA input-referred thermal-noise voltage is stated as for Fig \ref{fig5}.

\begin{multline}
S_{VIN}(thermal)=4kT[2\times \dfrac{(n\Gamma)_{17}}{g_{m17}}+2\times \dfrac{(n\Gamma)_{16}g_{m16}}{g{\textsuperscript 2}_{m17}}
+2\times \dfrac{(n\Gamma)_{11}g_{m11}}{g{\textsuperscript 2}_{m17}}+p\times \dfrac{(n\Gamma)_{14}g_{m14}}{g{\textsuperscript 2}_{m17}}\\
+2\times \dfrac{(n\Gamma)_{26}}{g_{m26}}
+2\times \dfrac{(n\Gamma)_{22}g_{m22}}{g{\textsuperscript 2}_{m26}}+2\times \dfrac{(n\Gamma)_{23}g_{m23}}{g{\textsuperscript 2}_{m26}}+2\times \dfrac{(n\Gamma)_{25}g_{m25}}{g{\textsuperscript 2}_{m26}}
+\dfrac{(n\Gamma)_{29}}{g_{m29}}+\dfrac{(n\Gamma)_{30}}{g_{m30}}+\dfrac{1}{R_{L}}]
\label{eqn:11}
\end{multline}

The description of the various parameters used in \ref{eqn:11} can be found in Table \ref{tbl2}.

\begin{table}[h!]
	\centering
	\renewcommand{\tabcolsep}{1mm}
	\caption{Description of various parameters used in \ref{eqn:11}}\label{tbl2}
	\begin{tabular}{l|l|l}
	\hline
		\bfseries Block& \bfseries Parameter & \bfseries Description \\
		\hline
		& k & Boltzman's constant \\
		& T & Temperature in Kelvin \\
		& n & Substrate factor of MOS transistor \\
		& $\Gamma$  & Thermal noise factor of MOS transistor\\
		\hline
		PTBTA & g{$_{\scriptsize m17}$} & Transconductance of transistors M17 and M18\\
		& g{$_{\scriptsize m16}$} & Transconductance of transistors M16 and M20 \\
		&g{$_{\scriptsize m11}$} & Transconductance of transistors M10 and M11 \\
		&g{$_{\scriptsize m14}$}&Transconductance of transistors M14 and M19 \\
		\hline
		TA &g{$_{\scriptsize m26}$} &  Transconductance of transistors M26 and M27 \\
		&g{$_{\scriptsize m22}$} & Transconductance of transistors M21 and M22 \\
		&g{$_{\scriptsize m23}$} & Transconductance of transistors M23 and M24 \\
		&g{$_{\scriptsize m25}$} & Transconductance of transistors M25 and M28  \\
		\hline
		I-V&g{$_{\scriptsize m29}$} & Transconductance of transistor M29\\
		converter&g{$_{\scriptsize m30}$} & Transconductance of transistor M30\\
		&R{$_{\scriptsize L}$} &  Load impedance \\
		\hline
	\end{tabular}
\end{table}

Due to the presence of current mirror devices along the entire signal channel, the factors 2 and p present in the noise expression. The first term in Eq.\ref{eqn:11} for the PTBTA portion of the readout circuit is gate-referred thermal-noise PSD of the input differential pair devices. The second, third, and fourth terms are equivalent to the current-mirror devices' drain-referred thermal-noise PSD. The TA component is denoted by the fifth, sixth, seventh, and eighth terms. The noise from the basic I-V converter is present in the ninth, tenth, and final terms.

Since the square of thermal noise (g{\textsuperscript 2}{$_{\scriptsize m17}$} or g{\textsuperscript 2}{$_{\scriptsize m26}$}) is inversely proportional to noise components, as shown in \ref{eqn:11}, a large value of {g{$_{\scriptsize m}$}} of input pair devices reduce thermal noise. Due to the increase in the effective value of {g{$_{\scriptsize m}$}}, the {g{$_{\scriptsize m}$}} boosting approach is advantageous to reduce the thermal noise contribution of the PTIA circuit while operating non-input devices at moderate inversion for moderate {g{$_{\scriptsize m}$}}. 

\begin{figure}[!h]
	\centerline{\includegraphics[scale=0.6]{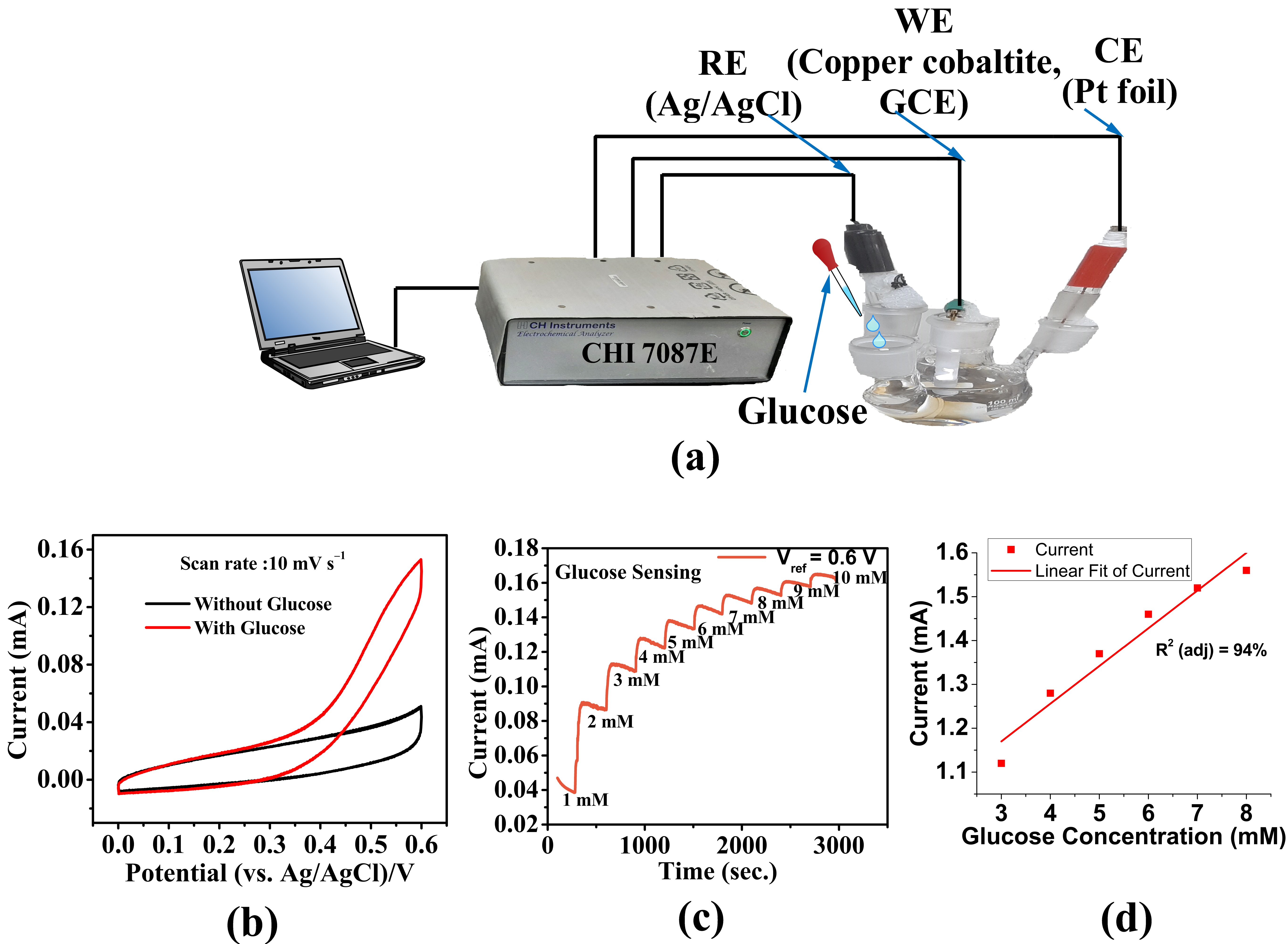}}
	\caption{Glucose Measurement (a) Glucose sensing setup (b) CV plot with and without glucose at scan rate of 10 mV/s (c) CA plot@0.6 V reference potential (d) Regression Fit relation between current and glucose concentration.}
	\label{fig6}
\end{figure} 

\section{Results and Discussion}
\label{sec3}
Using the simulation programme Cadence Virtuoso and the 0.18 $\mu$m CMOS process technology, the suggested circuit architecture for glucose level measurement is evaluated. For all simulations, the supply voltage V{$_{\scriptsize dd}$} = 1.5 V is used. The biassing currents for differential CC and TA's circuit are I{$_{\scriptsize b1}$} = 175 $\mu$A and I{$_{\scriptsize b2}$} = 350 $\mu$A, respectively. To achieve linear and efficient sensor current amplification, the resistance R{$_{\scriptsize z}$} was adjusted to 5 k$\Omega$. The following subsections \ref{sec3.1}, \ref{sec3.2}, \ref{sec3.3} and \ref{sec3.4}, respectively, discuss the electrochemical glucose measurement, results obtained for the proposed electronic glucose sensing architecture, measured results from emulated circuits, and comparison analysis of the proposed work.

\subsection{Electrochemical measurement details}\label{sec3.1}
The FT-Raman estimation was performed in an AIRIX STR 500 confocal Raman spectrometer equipped with a He-Cd laser, with an excitation wavelength of 532 nm and a spectral resolution of 1 cm{\textsuperscript {-1}}. The field emission scanning electron microscopy (FESEM) investigations using the FEI Nova Nano SEM 450 device was carried out. Prior to exposure to FESEM measurements, the material was uniformly distributed on carbon tape and finely crushed before being put on the sample holder.  

Ag/AgCl serves as the reference electrode in this three-electrode setup, while Pt foil (1$\times$2 cm) serves as the counter electrode. O{$_{\scriptsize 2}$} soaked 0.15 M aqueous NaOH solution served as the electrolyte for all electrochemical studies. The ultrasonic dissemination of 1.5 mg of the CuO/Cu{$_{\scriptsize 0.76}$}CO{$_{\scriptsize 2.25}$}O{$_{\scriptsize 4}$} material produced a homogeneous solution of 550 $\mu$L of deionized water and 50 $\mu$L of nafion solution (5 wt \% in ethanol), which served as the electrocatalyst or working electrode. A steady 15 muL drop of this homogenous solution was projected onto a clean GCE with a diameter of 5 mm, followed by drying at ambient temperature. By using amperometric current versus time measurements at 0.6 V vs Ag/AgCl and cyclic voltammetry (CV) at a specified scan rate of 10 mVs{\textsuperscript {-1}}, precise electrochemical estimates for glucose measurement were carried out. All of the electrochemical measurement studies were carried out using a CHI 7087E electrochemical workstation. The whole glucose electrochemical sensor setup, the CV, CA plot, and the relationship between current and glucose concentration are shown in Fig. \ref{fig6}. The sensor set-up described above has been used to monitor glucose concentrations in the range of 1 to 10 mM.

\subsection{Proposed electronics interface results}\label{sec3.2}
\begin{figure}[!h]
	\centerline{\includegraphics[scale=0.35]{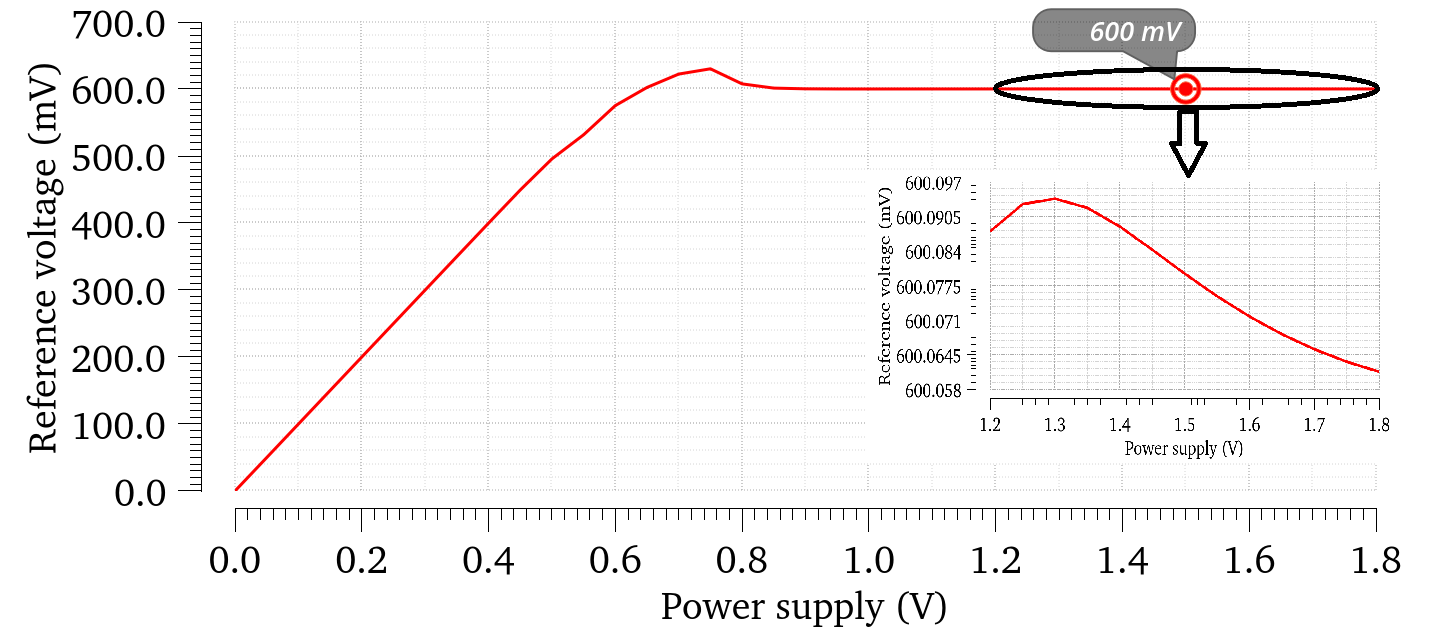}}
	\caption{0.6 V reference voltage produced by band gap voltage reference potentiostat circuit.}
	\label{fig7}
\end{figure}

\begin{figure}[!h]
	\centerline{\includegraphics[scale=0.8]{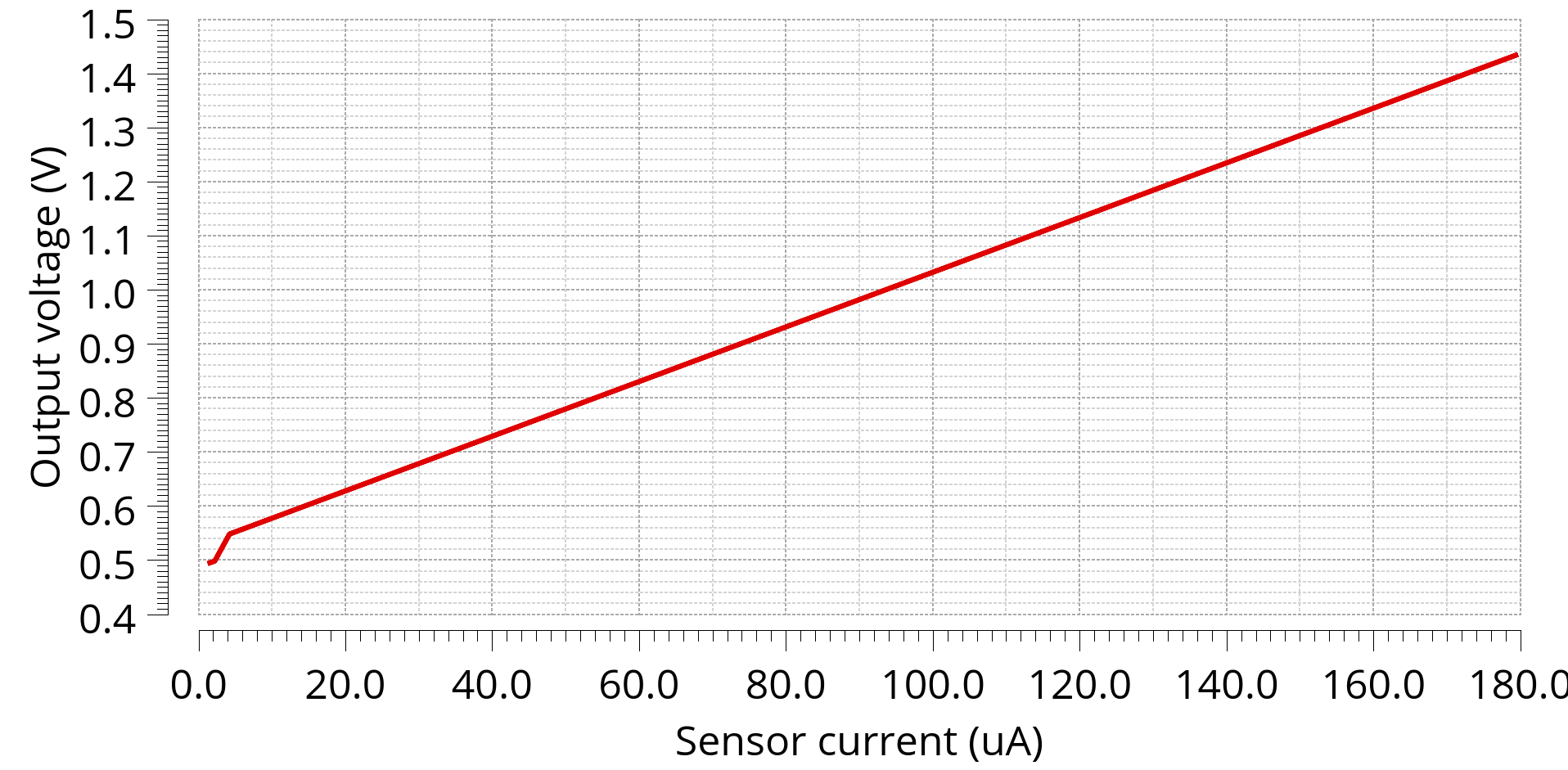}}
	\caption{Output voltage and sensor current ralation of readout circuit.}
	\label{fig8}
\end{figure}

\begin{figure}[!h]
	\centerline{\includegraphics[scale=0.45]{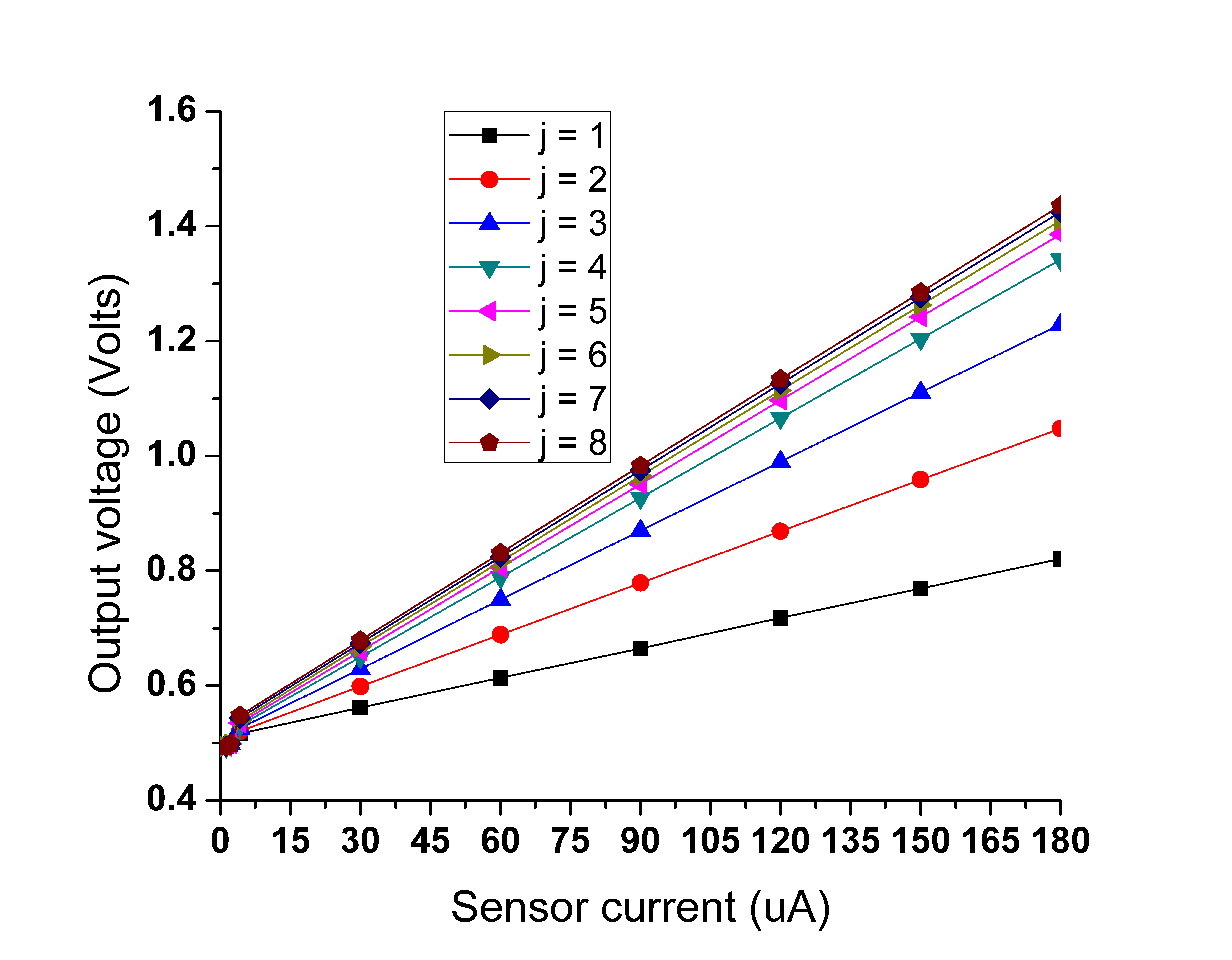}}
	\caption{Programmable output voltage and sensor current relation of readout circuit.}
	\label{fig9}
\end{figure}

The bandgap reference voltage potentiostat's performance behaviour is depicted in Fig. \ref{fig7}, and it can be seen that it maintains 0.6 V despite a $\pm$20\% supply voltage change. Bandgap reference voltage potentiostat circuit implementation (Fig. \ref{fig2}). According to Fig. \ref{fig8}, the PTIA's linear output voltage ranges from 0.55 V to 1.44 V while its current variation extends from 4.2 $\mu$A to 180 $\mu$A. Fig. \ref{fig9} depicts the PTIA circuit's customizable output voltage and sensor current relationship.


\begin{figure}[!h]
	\centerline{\includegraphics[scale=0.9]{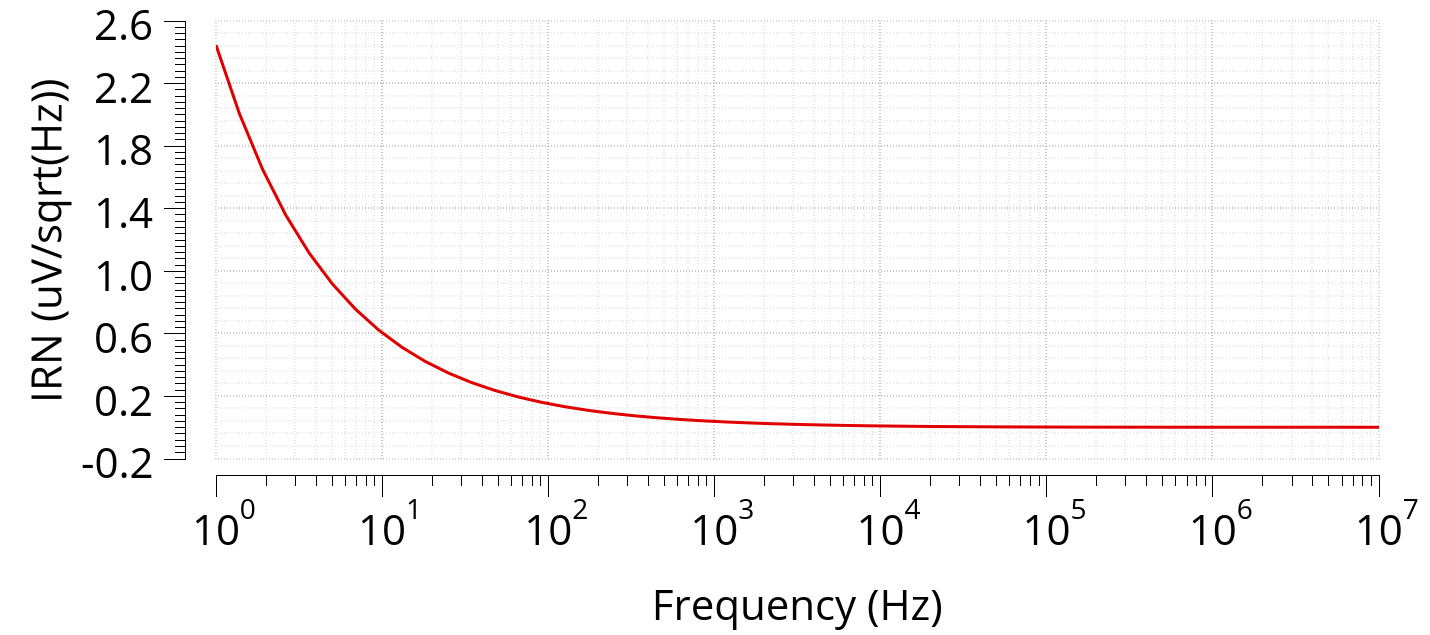}}
	\caption{Input referred noise of proposed readout circuit.}
	\label{fig10}
\end{figure}

\begin{figure}[!h]
	\centerline{\includegraphics[scale=0.78]{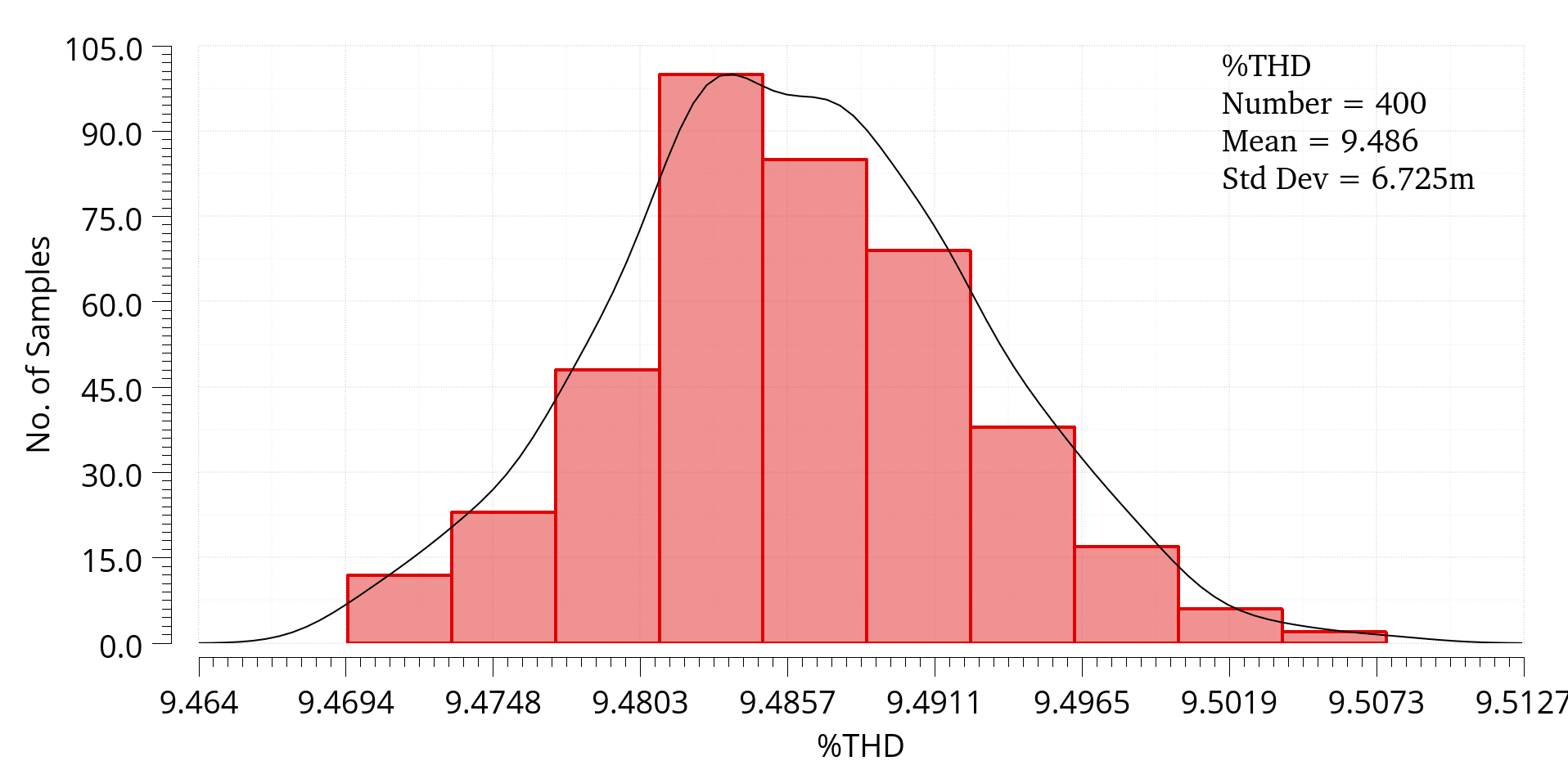}}
	\caption{Monte Carlo analysis of \%THD.}
	\label{fig11}
\end{figure}

Fig. \ref{fig10} displays the input referred noise (IRN) response of the suggested design. Its {V{$_{\scriptsize rms}$}} value for frequencies between 1 Hz and 10 kHz is 5.101 $\mu${V{$_{\scriptsize rms}$}}.
The projected \%THD ranges from 7.6 to 10.2 for the current fluctuation of 4.2 $\mu$A to 180 $\mu$A, which is within acceptable bounds. To test the design's resilience to transistor mismatch and process changes, the Monte Carlo analysis was done for \%THD. As shown in Fig. \ref{fig11}, the mean and standard deviations for the 400 samples are 9.486 \% and 0.00067 \%, respectively.

\begin{figure}[!h]
	\setlength{\fboxrule}{1pt}
	\centerline{\fbox{\includegraphics[width=8.5cm, height=8.5cm]{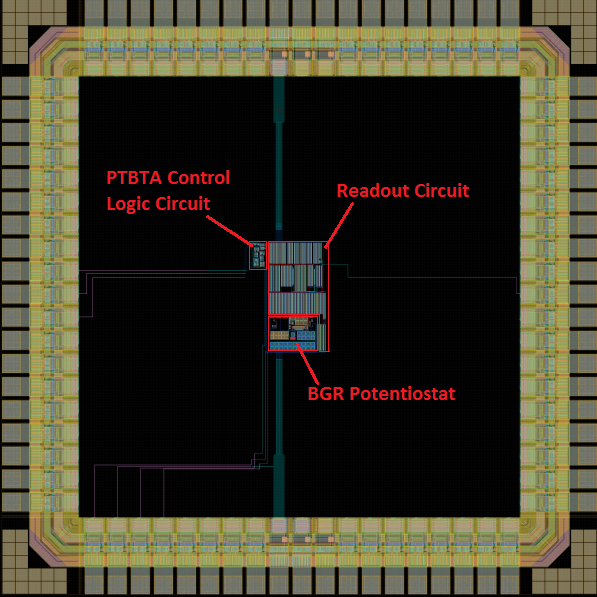}}}
	\caption{Layout of the IC in 0.18 $\mu$m technology of proposed electronic interface.}
	\label{fig12}
\end{figure}

Tables \ref{tbl3}, \ref{tbl4} and \ref{tbl5}, respectively, show the impacts of process, voltage, and temperature variation on key parameters of suggested glucose sensor structures. The layout is depicted in Fig. \ref{fig12} has the core area 0.0684 mm{\textsuperscript 2} (190 $\mu$m x 360 $\mu$m). The output voltage and sensor current relationship post-layout simulation results differ by 8.5\% from pre-layout simulation results.

\begin{table}[h!]
	\centering
	\renewcommand{\tabcolsep}{2.5mm}
	\caption{Effect of process variations on the performance of proposed glucose sensing architecture}\label{tbl3}
	\begin{tabular}{l l l l l l} 
		\hline
		\multicolumn{1}{l}{{{\bfseries Parameters}}} & \multicolumn{5}{l}{{{\bfseries Process variations (V{$_{\scriptsize dd}$} = 1.5 V, temp = 27 \textdegree C )}}}\\
		\cline{2-6}
		& ss & sf &  tt & fs & ff \\
		\hline
		Output 	voltage (V)  & 1.03& 1.031&1.033 &1.034 &1.037\\
		IRN ($\mu$V) &2.442&2.443&2.446&2.449&2.452\\
		\%THD &9.472&9.489&9.486&9.491&9.498\\
		\hline
	\end{tabular}
\end{table}
\begin{table}[h!]
	\centering
	\renewcommand{\tabcolsep}{2mm}
	\caption{Effect of voltage variations on the performance of proposed glucose sensing architecture}\label{tbl4}
	\begin{tabular}{l l l l l l} 
		\hline
		\multicolumn{1}{l}{{{\bfseries Parameters}}} & \multicolumn{5}{l}{{{\bfseries Voltage variations (Process = tt, temp = 27 \textdegree C )}}}\\
		\cline{2-6}
		& 1.35 V & 1.425 V &  1.5 V & 1.575 V& 1.65 V \\
		\hline
		Output 	voltage (V)  & 0.896& 0.96&1.033 &1.127 &1.237\\
		IRN ($\mu$V) &2.426&2.435&2.446&2.458&2.471\\
		\%THD &13.12&11.06&9.486&8.476&7.934\\
		\hline
	\end{tabular}
\end{table}
\begin{table}[h!]
	\centering
	\renewcommand{\tabcolsep}{1mm}
	\caption{Effect of temperature variations on the performance of proposed glucose sensing architecture}\label{tbl5}
	\begin{tabular}{l l l l l l l l l} 
		\hline
		\multicolumn{1}{l}{{{\bfseries Parameters}}} & \multicolumn{8}{l}{{{\bfseries Temperature variations (V{$_{\scriptsize dd}$} = 1.5 V, process = tt)}}}\\
		\cline{2-9}
		& -40\textdegree C & -20\textdegree C &  0\textdegree C & 20\textdegree C & 40\textdegree C & 60\textdegree C&  80\textdegree C& 100\textdegree C\\
		\hline
		Output  & 1.079& 1.053&1.04 &1.034 &1.032 &1.036& 1.05&1.097\\
		voltage (V)&&&&&&&&\\
		IRN ($\mu$V) &2.27&2.32&2.38&2.43&2.48&2.53&2.58&2.63\\
		\%THD &8.91&8.86&8.72&9.21&10.12&11.32&12.7&14.17\\
		\hline
	\end{tabular}
\end{table}

\subsection{Experimental setup for glucose concentration measurement}\label{sec3.3}
\begin{figure}[!h]
	\centerline{\includegraphics[scale=0.55]{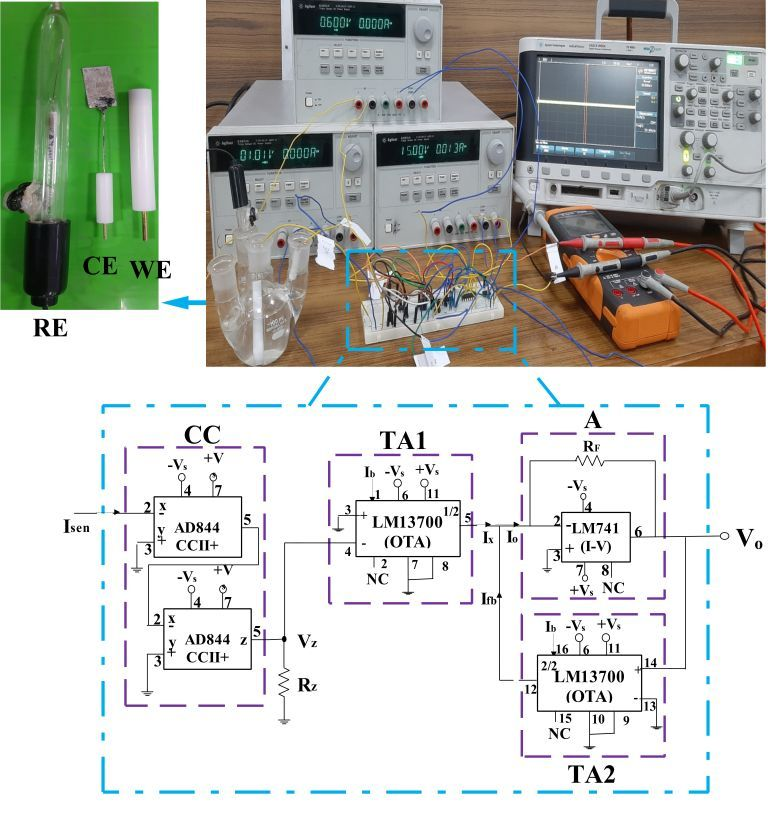}}
	\caption{Experimental setup for glucose concentration measurement with emulated readout circuit.}
	\label{fig13}
\end{figure}

\begin{figure}[!h]
	\centerline{\includegraphics[scale=0.38]{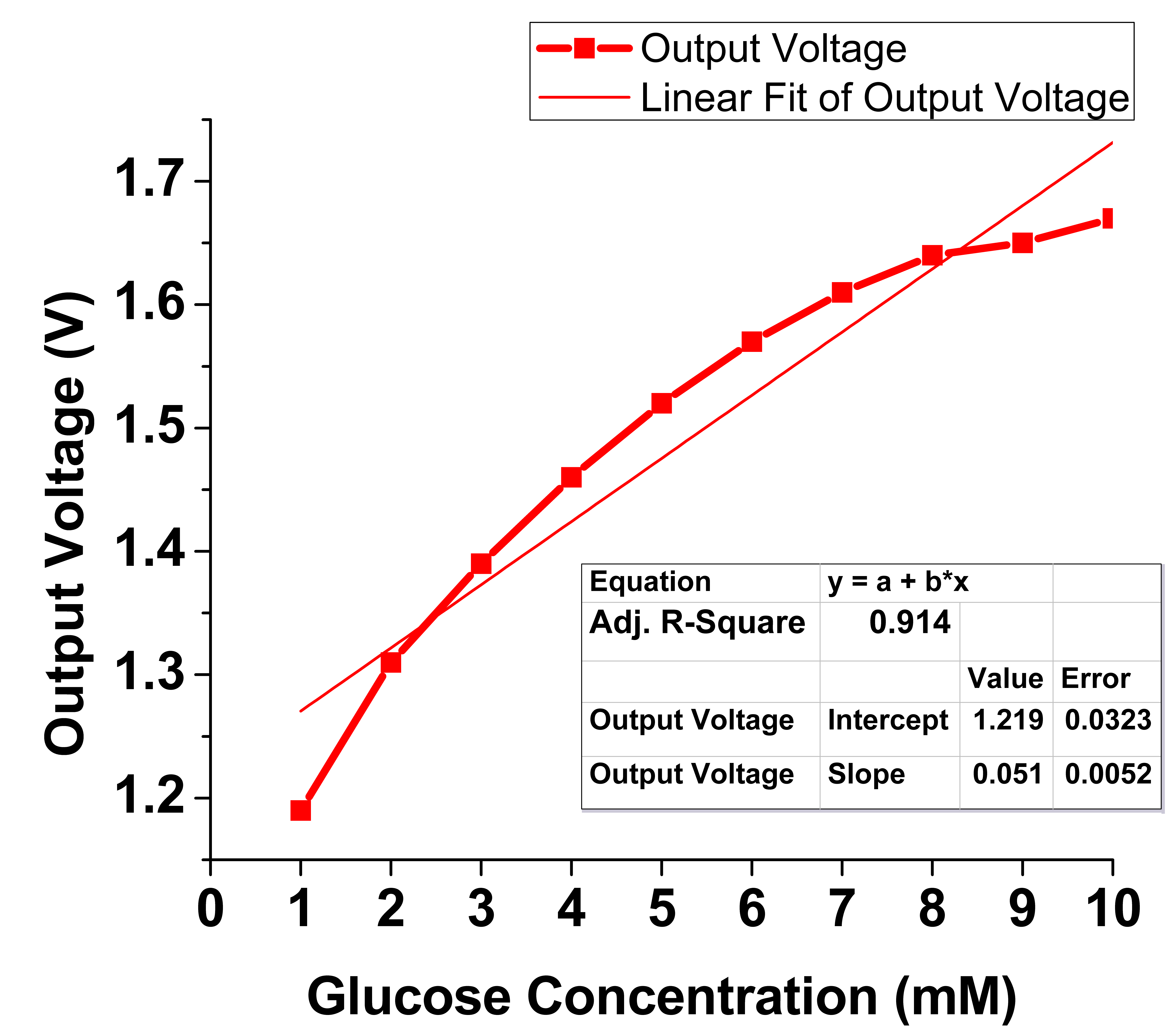}}
	\caption{Measured output voltage and glucose concentration relation by experimental setup for glucose sensing.}
	\label{fig14}
\end{figure}

In Fig. \ref{fig13}, the entire setup was simulated for glucose measurement. Two AD844 integrated circuits are used to implement the current conveyor of the second generation (CCII-). The IC LM13700 is used to implement the two TAs, and the op-amp IC LM741 and feedback resistance are used to accomplish the I-V conversion. Consideration is given to the 0.6 V reference voltage. The proposed block design for glucose sensing is strengthened by the measurement setup, which demonstrates its practical viability. After a predetermined amount of time, the glucose was added to the four arms of the beaker in steps of 1 mM, and measurements were taken up to 10 mM. As indicated in Fig. \ref{fig14}, voltage measurements were made in the 1.19 V to 1.67 V range, which corresponds to glucose concentration ranges of 1 mM to 10 mM.


\subsection{Comparison}\label{sec3.4}
Table \ref{tbl6} compares the performance of the suggested design with the existing literature. \cite{martin2009fully,shenoy2014,ghore2017} reported significant power dissipation, while \cite{shenoy2014,ghore2017,al2016} provided a comparatively modest transimpedance gain. It was noted in \cite{martin2009fully,shenoy2014} that IRN had a higher value. The literature \cite{ahmadi2008current} has a small dynamic range but a low power dissipation. In contrast to the suggested design, the article \cite{al2016} has a low power dissipation but gives a limited range of glucose concentration detection. When compared to \cite{martin2009fully,ghore2017} the core area of a complete architecture is low enough. According to the findings, the proposed design is the only one that has programmable properties, a large dynamic range, low power dissipation, and a wide range of glucose detection. 

\begin{table}[t]\scriptsize
	\centering
	\renewcommand{\tabcolsep}{1mm}
	\caption{Comparison of proposed design with existing literatures}\label{tbl6}
	\begin{tabular}{l l l l l l l}
		\hline
		\bfseries Parameter &\bfseries \cite{ahmadi2008current}& \bfseries \cite{martin2009fully} & \bfseries \cite{shenoy2014} & \bfseries \cite{ghore2017} &\bfseries \cite{al2016}& \bfseries This work \\
		\hline
		Process ($\mu$m)&0.18&0.18&0.35&0.35&0.18&0.18\\
		CE&Pt*&Pt wire&Pt/MWCNTs/&Pt&Ag/AgCl&Pt foil\\
		&&&microneedle&&&\\
		WE&Oxidzed hydrogen&Pt&Pt/MWCNTs/&Pt-nanostructured&Vertically aligned carbon&CuO/CuO.76CO2.25O4\\
		&peroxide*&&microneedle&consisting of glucose oxidase&nanofiber microelectrode&coated GCE filled\\
		RE&Ag/AgCl*&Calomel&SCE-calomel&Ag/AgCl&Ag/AgCl&Ag/AgCl\\
		Power supply&1.8&1.8&1.65&3.3&1.8&1.5\\
		Current supply&38.89$\mu$A&8.8 mA&5$\mu$A&2.82 mA&39.83$\mu$A&350 $\mu$A\\
		Power dissipation&70$\mu$W&15.84 mW&5.1 mW &9.3 mW&71.7 $\mu$W&2.33 mW\\
		Gain&NA&NA&44.3 k$\Omega$&42 k$\Omega$&39.8 k$\Omega$&17.3-50.5 k$\Omega$\\
		IRN(/\(\sqrt{Hz}\))&NA&2.5 mV&6.6 mV&0.47 pA&0.69 $\mu$A&2.46 $\mu$V\\
		Sensor current&1n-1$\mu$&NA&5-30 $\mu$A&-20-20$\mu$A&500 nA-7 $\mu$A&4.2-180$\mu$A\\
		detection range&& &&&\\
		DR (dB)&60&63&46.1&156&150.7&131.4\\
		Prog. feature&No&No&No&No&No&Yes\\
		Glucose conc. range&0-40 mM&1-10 mM&2.5-10 mM&100$\mu$M-1 mM&0.4-4 mM&1-10 mM\\
		Core area (mm{\textsuperscript 2})&0.02&0.45&0.0501&0.6&0.0179&0.0684\\
		\hline
	\end{tabular}
\end{table}

\section{Conclusion} \label{sec4}
In this research, a novel electronics interface architecture for glucose sensing applications is provided, along with measured findings from an emulated circuit. The performance results are validated using the 0.18 $\mu$m, CMOS process technology by the Cadence Virtuoso. The proposed design has been supported by a number of simulations. The measurement system can identify glucose concentrations between 1 and 10 mM. The suggested glucose interface architecture generates an output voltage that is linearly spaced between 0.55 V and 1.44 V for input sensor current ranges between 4.2 $\mu$A and 180 $\mu$A. A high dynamic range of 131.4 dB is attained, with a THD of roughly 7.6–10.2\%. The suggested design can be useful for the detection of additional target objects utilised in electrochemical sensing applications, such as uric acid, tears, and so on. 

\section*{Acknowledgment}
The VLSI Lab in the ECE Department of MNIT Jaipur and the Materials Electrochemistry \& Energy Storage Lab in the Chemistry Department of MNIT Jaipur provided the support of EDA tools and measurement setup to carry out the experiments, respectively, and the authors are appreciative. 

\bibliographystyle{IEEEtran}
\bibliography{Ref}

\pagebreak
\section*{About the Authors}
\begin{IEEEbiography}[{\includegraphics[width=1in,height=1.25in,clip,keepaspectratio]{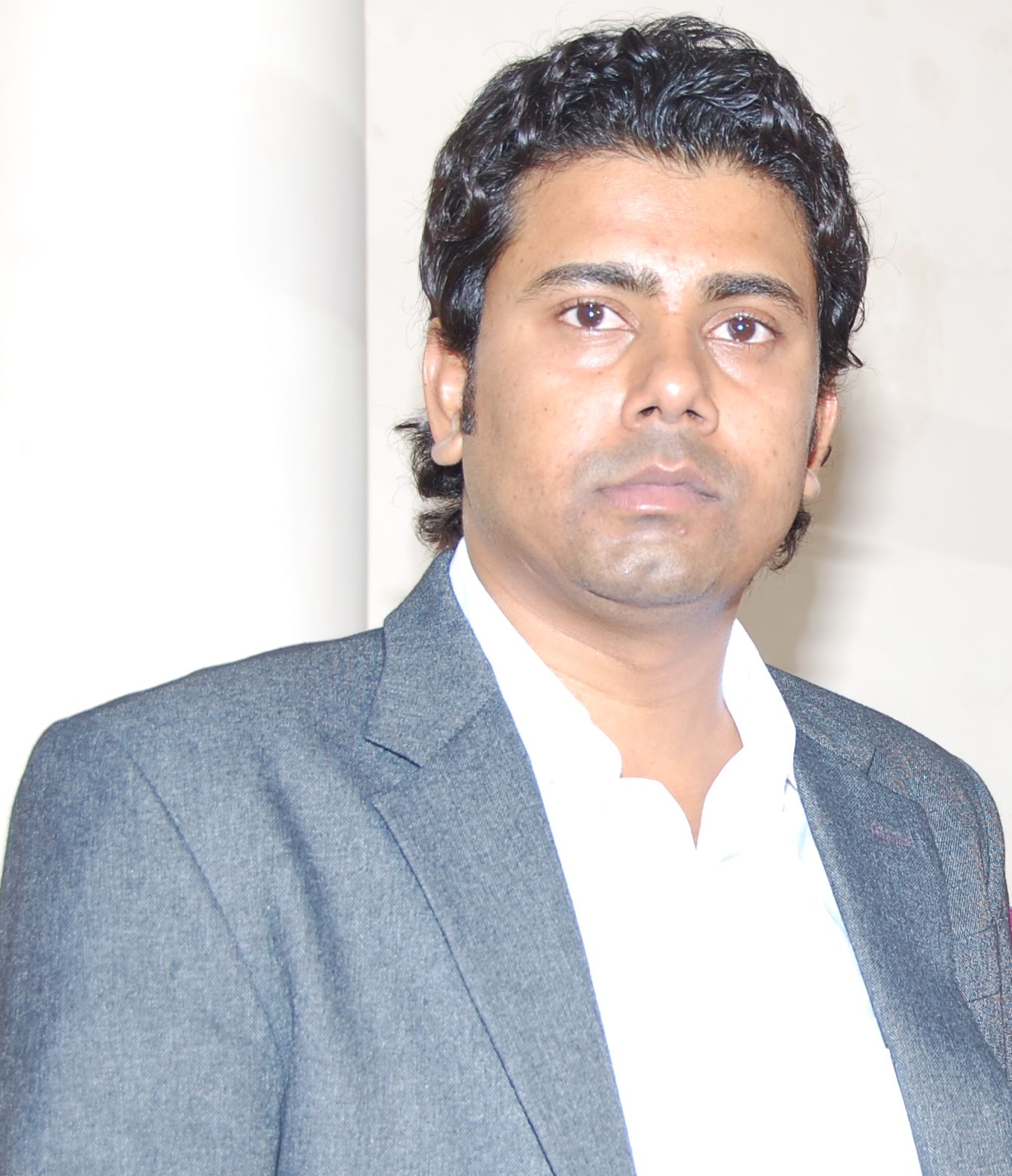}}]{Riyaz Ahmad} (Member, IEEE) received the M.Tech degree in VLSI and Embedded System from Sardar Vallabhbhai National Institute of Technology (SVNIT) Surat, India in 2014. During 2008-2011, he worked as Lecturer in Galgotias College of Engg. and Technology Greater Noida, India. He was teaching assistant at Sardar Vallabhbhai National Institute of Technology (SVNIT) Surat, India during 2015-2017. He worked as Assistant Professor (Under NPIU, TEQIP-III Project of MHRD, Govt. of India and World Bank) at University Departments, Rajasthan Technical University Kota, Rajasthan, from 2018 to 2021. He is currently working towards PhD degree and also working as JRF for Reagional Academic Center for Space sponsered research project , at Malaviya National Institute of Technology (MNIT) Jaipur, India, where his research focuses on design of circuits in Analog VLSI for biomedical applications. He has thorough experience on working with various industry standard VLSI design tools (Cadence Virtuoso, Mentor Graphics, Tanner EDA).  He has published article in various journals and conferences.
\end{IEEEbiography}

\vspace{-0.7cm}

\begin{IEEEbiography}[{\includegraphics[width=1in,height=1.25in,clip,keepaspectratio]{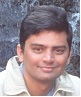}}]{Amit mahesh Joshi} (Member, IEEE) received the Ph.D. degree from the NIT, Surat, India. He is currently an Assistant Professor at National Institute of Technology, Jaipur. His area of specialization is Biomedical signal processing, Smart healthcare, VLSI DSP Systems and embedded system design. He has also published papers in international peer reviewed journals with high impact factors. He has published six book chapters and also published more than 70 research articles in excellent peer reviewed international journals/conferences. He has worked as a reviewer of technical journals such as IEEE Transactions/ IEEE Access, Springer, Elsevier and also served as Technical Programme Committee member for IEEE conferences which are related to biomedical field. He also received honour of UGC Travel fellowship, the award of SERB DST Travel grant and CSIR fellowship to attend well known IEEE Conferences TENCON, ISCAS, MENACOMM etc across the world. He has served as session chair at various IEEE Conferences like TENCON -2016, iSES-2018, iSES-2019, ICCIC-14 etc. He has also supervised 19 PG Dissertations and 16 UG projects. 
He has completed supervision of 4 Ph.D thesis and six more research scholars are also working. 
\end{IEEEbiography}

\vspace{-0.7cm}

\begin{IEEEbiography}[{\includegraphics[width=1in,height=1.25in,clip,keepaspectratio]{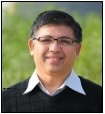}}]{Dharmendar Boolchandani} (Member, IEEE)  received Bachelor’s degree in Electronics from MREC, Jaipur in 1988 and M.Tech. from CEDT, IISc Bangalore in 1998. He obtained his Ph.D. from MNIT, Jaipur in 2011. He is currently Professor in the Department of ECE, MNIT, Jaipur, India. His current research areas include Macromodeling for Analog subsystems and interconnect, sensor interfacing circuits, MEMS based inertial sensors.
\end{IEEEbiography}

\end{document}